\documentclass[11pt,twoside]{article}
\usepackage{./asp2014}

\aspSuppressVolSlug
\resetcounters

\begin{document}

\title{Interstellar medium, young stars, and astrometric binaries in Galactic archaeology spectroscopic surveys}
\author{Toma\v{z} Zwitter,$^1$ Janez Kos,$^1$ Maru\v{s}a \v{Z}erjal,$^1$ and Gregor Traven$^1$
\affil{$^1$University of Ljubljana, Faculty of Mathematics and Physics, Ljubljana, Slovenia; \email{tomaz.zwitter@fmf.uni-lj.si}}}

\paperauthor{Toma\v{z} Zwitter}{tomaz.zwitter@fmf.uni-lj.si}{}{University of Ljubljana}{Faculty of Mathematics and Physics}{Ljubljana}{}{1000}{Slovenia}
\paperauthor{Janez Kos}{janez.kos@fmf.uni-lj.si}{}{University of Ljubljana}{Faculty of Mathematics and Physics}{Ljubljana}{}{1000}{Slovenia}
\paperauthor{Maru\v{s}a \v{Z}erjal}{marusa.zerjal@fmf.uni-lj.si}{}{University of Ljubljana}{Faculty of Mathematics and Physics}{Ljubljana}{}{1000}{Slovenia}
\paperauthor{Gregor Traven}{gregor.traven@fmf.uni-lj.si}{}{University of Ljubljana}{Faculty of Mathematics and Physics}{Ljubljana}{}{1000}{Slovenia}

\begin{abstract}
Current ongoing stellar spectroscopic surveys (RAVE, GALAH, Gaia-ESO, LAMOST, APOGEE, Gaia) are mostly devoted to studying Galactic archaeology and structure of the Galaxy. But they allow for important auxiliary science: (i) Galactic interstellar medium can be studied in four dimensions (position in space $+$ radial velocity) through weak but numerous diffuse insterstellar bands and atomic absorptions seen in spectra of background stars, (ii) emission spectra which are quite frequent even in field stars can serve as a good indicator of their youth, pointing e.g. to stars recently ejected from young stellar environments, (iii) astrometric solution of the photocenter of a binary to be obtained by Gaia can yield accurate masses when joined by spectroscopic information obtained serendipitously during a survey.

These points are illustrated by first results from the first three surveys mentioned above. These hint at the near future: spectroscopic studies of the dynamics of the  interstellar medium can identify and quantify Galactic fountains which may sustain star formation in the disk by entraining fresh gas from the halo; RAVE already provided a list of $\sim$~14,000 field stars with chromosperic emission in Ca II lines, to be supplemented by many more observations by Gaia in the same band, and by GALAH and Gaia-ESO observations of Balmer lines; several millions of astrometric binaries with periods up to a few years which are being observed by Gaia can yield accurate masses when supplemented with measurements from only a few high-quality ground based spectra.
\end{abstract}

\section{Introduction}

Last decade has seen a major shift in stellar spectroscopy: a slow collection of individual spectra has been accelerated by massive surveys, mostly using fiber-fed spectrographs with hundreds of spectra observed simultaneously. The past and ongoing efforts include RAVE  \citep{Steinmetz2006,Zwitter2008,Siebert2011,Kordopatis2013}, Gaia-ESO \citep{Gilmore2012}, SEGUE \citep{Yanny2009}, APOGEE \citep{Zasowski2013}, LAMOST \citep{Luo2015}, GALAH \citep{DeSilva2015}, and of course Gaia \citep{Prusti2014}. Up-to-date overviews of the state and results of these surveys are given elsewhere in this volume. 

The main goal of stellar spectroscopic surveys is to study Galactic structure and evolution. But the collected spectra allow for a significant auxiliary science. The three examples discussed below are an illustration of a vast range of posibilities and are by no means exhaustive. We believe that every observer could add further relevant uses of hundreds of thousands of stellar spectra, which were in most cases selected for observation only following simple positional and magnitude constraints. The first example illustrates research of the multi-dimensional structure of the interstellar medium. The next one helps with identifying young stars in the field. The last one is an example on how even a single spectrum obtained by a stellar survey can improve the solution of an astrometric binary which is being derived by Gaia. 

\section{Interstellar Medium}

In 2020, the Gaia mission (launched in December 2013) is expected to release 6-dimensional (spatial position + velocity) vectors for a significant fraction of stars on our side of the Galactic centre, thus allowing a computation of stellar orbits and of evolution of the Galaxy as a whole. Traditional studies of the Galactic interstellar medium (ISM) cannot yield information equivalent to stars, as absorption studies get only a 2-dimensional (column density) information by observing one hot star at a time. But ISM allows to open up its 3-rd and 4-th dimension by studying diffuse interstellar bands (DIBs), weak but numerous absorption lines seen in spectra of background stars which are likely caused by distinct macromolecular carriers. High dimensionality requires measurement of the strength of these weak interstellar lines also for cool stars which by far outnumber hot stars in the Galaxy. Recent new approaches divide out the cool star spectrum by use of synthetic models of stellar atmospheres \citep{Puspitarini2015} or in a self-calibrated way by using spectra of similar stars with negligible ISM absorption observed at high Galactic latitudes by the same survey \citep{Kos2013}. By observing a given DIB toward many stars which are nearly in the same direction but at different and known distances one can reconstruct absorption sites along the line of sight. Joining observations in many directions on the sky then gives their spatial distribution. Finally, measurement of radial velocity shift yields a 4-dimensional picture of the ISM for each DIB, and can even constrain placement of multiple clouds along each line of sight. Interstellar absorption lines of sodium and potassium atoms yield information equivalent to DIBs, but emission lines or dust absorptions are limited to up to 3 dimensions.

ISM is the place of violent collisions of supernova shells, plus winds from asymptotic giant branch stars and hot-star associations. Head-on collisions in the Galactic plane are difficult to interpret, though an expected Galactic rotation pattern has been nicely identified \citep{Zasowski2015}. But observations of the on-going GALAH and partly Gaia-ESO surveys are away from the plane where interactions generally result in a net motion perpendicular to the plane. If any shells of absorbing material are identified we can assume that their motion is perpendicular to shell surfaces and reconstruct a complete velocity vector from its radial velocity component. Such information for ISM is then equivalent to the one collected for stars by Gaia.

This information can be used to study past events in the interstellar medium. \citet{Kos2014} published a quasi 3-dimensional map of intensity of diffuse interstellar band at 8620~\AA\ which shows that distribution of DIB extinction is thicker than the one of dust and that it is different on either side of the Galactic plane, a witness to asymmetries in placement of recent explosions of supernovae and to incomplete vertical mixing. Observations with the Gaia-ESO and GALAH surveys could be used to 
increase the dimensionality of ISM studies to 4 dimensions 
\citep[for an example of radial velocity measurements see][]{Kos2015}. They could also identify and characterize Galactic fountains blown away by supernovae in the last million years. Such flows are thought to sustain star formation in the disk by entraining fresh gas from the halo, so they provide a mechanism which explains why star formation in our and other similar galaxies did not stop when gas present in the disk has been used up \citep{BlandHawthorn2009,Fraternali2014}.


\articlefigure[width=\textwidth]{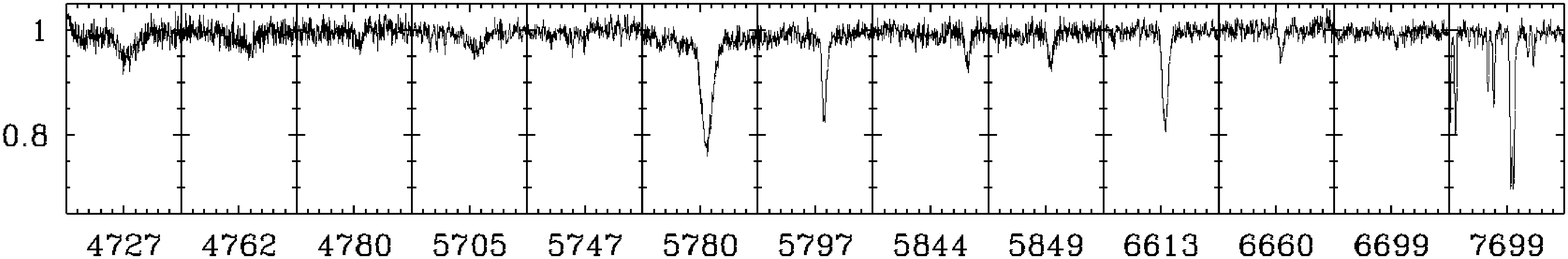}{figDIBsGALAH}{
Diffuse interstellar bands and the K~I interstellar atomic line at 7699\AA\ in GALAH spectra.
}

Figure \ref{figDIBsGALAH} plots a dozen DIBs and the K~I interstellar atomic line at 7699~\AA\ in a stellar spectrum observed by GALAH. Spectrum of TYC 4011-102-1, a hot star with strong interstellar absorptions close to the Galactic plane, is shown. Each 20~\AA\ wide panel is centred on the DIB wavelength as listed in \citet{Jenniskens1994}. Plotted wavelengths are heliocentric. Right-most panel identifies two interstellar clouds for K~I at different velocities. For a majority of GALAH objects, which lie away from the Galactic plane, such complications are rare (but can be detected). 

\section{Young stars in the field}

Properties of a star are entirely determined by its initial composition, mass and current age if one neglects rotation, magnetism or multiplicity. As noted by David \citet{Soderblom2014} "age is not a direct agent of change and cannot be measured like mass or composition. Also, age affects the core of the star, but we observe the surface which is complex." Large spectroscopic surveys have the possibility to measure some empirical age indicators, i.e.\ rotation, activity, and lithium depletion boundary. The GALAH survey will bring studies of these age indicators to industrial scale with its hundreds of thousands of observed spectra. Lithium depletion studies have been motivated by lithium observations of main-sequence stars in young clusters and the halo
\citep[e.g.\ ][]{Soderblom1995}. The GALAH survey includes the Li~I 6708~\AA\ line in its red channel. Its resolving power of $\sim 28,000$ and a typical S$/$N ratio of 100 per resolution element allow for efficient measurement of stellar rotation and for studies of profiles of H$\alpha$ and H$\beta$ lines which are sensitive to chromospheric activity. Measurement of these youth indicators for field stars is important, as it may point to stars recently ejected from young stellar environments. Parallaxes and proper motions measured by Gaia, together with spectroscopically derived radial velocities permit to reconstruct their Galactic orbits and so to identify recently dispersed stellar clusters. Multidimensional chemistry studies, which are within the scope of GALAH, are then the final check of the emerging picture based on chemical tagging \citep{Freeman2002}. 

\articlefigure[width=\textwidth]{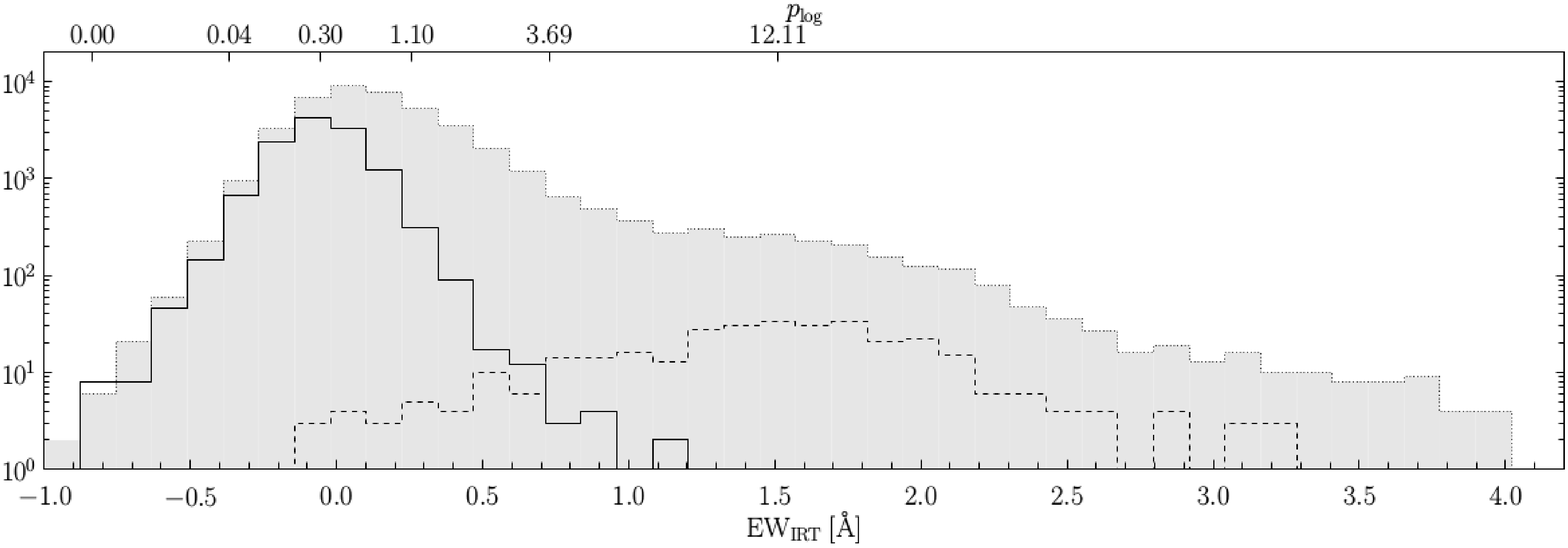}{figactivityRAVE}{
Distribution of the equivalent width of emission component of Calcium infrared triplet for active stars in RAVE (grey area). Solid histogram are normal stars which are assumed to be inactive, while dashed line marks pre-main sequence stars known to Simbad. From \citet{Zerjal2013}.}

Stellar activity identification is now entering the era of massive studies. 
Figure \ref{figactivityRAVE} summarizes active star candidates found in RAVE data using equivalent width of the emission components of the Ca~II infrared triplet lines \citep[EW$_{IRT}$, ][]{Zerjal2013}. Grey histogram is a distribution of EW$_{IRT}$ for 
stars with active morphology in RAVE, as identified by a locally linear embedding technique \citep{Matijevic2012}. Solid line marks normal stars which are assumed to be inactive, while dashed line marks RAVE stars classified by Simbad to be pre-main sequence stars. $p_{log}$ is a logarithmic measure for the probability that a star with a given EW$_{IRT}$ differs from an inactive spectrum. Its values from left to right correspond to the probabilities of 5 and 2~$\sigma$ below zero, zero, and 2, 5 and 10~$\sigma$ above zero. Altogether the work identifies $\sim 14,000$ stars with  chromospheric flux in Ca~II lines detectable at least at a 2~$\sigma$ confidence level.

\begin{table}[!ht]
\caption{Fractions of categorisations of the same object into subtypes (see text) for emission type objects in the Gaia-ESO survey. From \citet{Traven2015}.}
\smallskip
\label{tablefractions}
\begin{center}
\begin{tabular}{crrrrrrr}
Prevalent&\multicolumn{7}{c}{Categories of spectra of the same object (\%)}\\
category &$E_{bl}$&$E_{sp}$&$E_{dp}$&$P_{Cyg}$&$IP_{Cyg}$&$S_{abs}$&$E_{abs}$\\
\noalign{\smallskip}
\tableline
$E_{bl}$  &80.2& 0.8& 3.0& 0.4& 0.5&13.5& 1.5\\
$E_{sp}$  & 1.4&80.0& 0.1& 0.5& 4.0& 3.6&10.4\\
$E_{dp}$  & 4.0& 0.1&73.6& 1.7& 1.3&18.6& 0.6\\
$P_{Cyg}$ & 1.9& 9.7& 1.0&64.0& 3.5& 3.3&16.7\\
$IP_{Cyg}$& 1.4& 5.2& 0.1& 0.2&77.2& 5.3&10.5\\
$S_{abs}$ &12.1& 2.5&13.2& 1.5& 1.8&67.1& 1.9\\
$E_{abs}$ & 0.3& 4.3& 0.1& 1.7& 5.3& 1.6&86.7\\
\noalign{\smallskip}
\tableline
\end{tabular}
\end{center}
\end{table}

Presence of emission components in the Ca~II infrared triplet (RAVE, Gaia, and Gaia-ESO) or in Balmer lines (GALAH and Gaia-ESO) do not prove that the object is young: interacting binaries are an obvious example of old objects with emission type spectra. But such objects are not very common in the field. RAVE (Fig.\ \ref{figactivityRAVE}) found that strong emissions suggest a pre main-sequence evolutionary phase. This is consistent with results of the Gaia-ESO survey, where \citet{Traven2015} studied 22,035 spectra of stars in young open cluster fields and found that 7698 spectra (35\%) belonging to 3765 stars have intrinsic emission in H$\alpha$. Again, such a large fraction of emission type spectra in a young stellar environment suggests that emission is related to youth. But emission is a transient property and morphological classification of emission may be changing with time. \citet{Traven2015} shows that most profiles are composed and classifies such profiles by properties of fits using two Gaussians: $E_{bl}$ stands for blended emission components, $E_{sp}$ have double sharp peaks, $E_{dp}$ are double emission, $P_{Cyg}$ are P-Cygni profiles, $IP_{Cyg}$ are inverted P-Cygni, $S_{abs}$ is self-absorption and $E_{abs}$ is emission within absorption. Off-diagonal elements in table \ref{tablefractions} report correlations betweeen individual composed profile types. When emission blend is the prevalent category for an object, it is most often in combination with self-absorption, which is best explained by one of the two components being in transition between absorption and emission. Similarly, the double sharp peaks can change to emission with absorption if a relatively weak absorption is constantly present and one of the emission peaks diminishes. The largest off-diagonal elements connect double peaks and self-absorption. The distinction between the two categories is largely influenced by the inclination of the slopes in the profile that are liable to change in the presence of additional weaker components or they are harder to retrieve in the case of more noisy spectra. The most frequently identified morphological categories from \citet{Traven2015} are emission blend (1729 spectra), emission in absorption (1652 spectra), and self absorption (1253 spectra). We conclude that many stars have their emission transient in time or in morphological type, so that activity detected through emission is an indication of youth which is not always present and should be used in connection with the absolute position of the star on the H-R diagram, a frequently known property in the Gaia era. 

\section{Astrometric binaries}

\articlefigure[width=\textwidth]{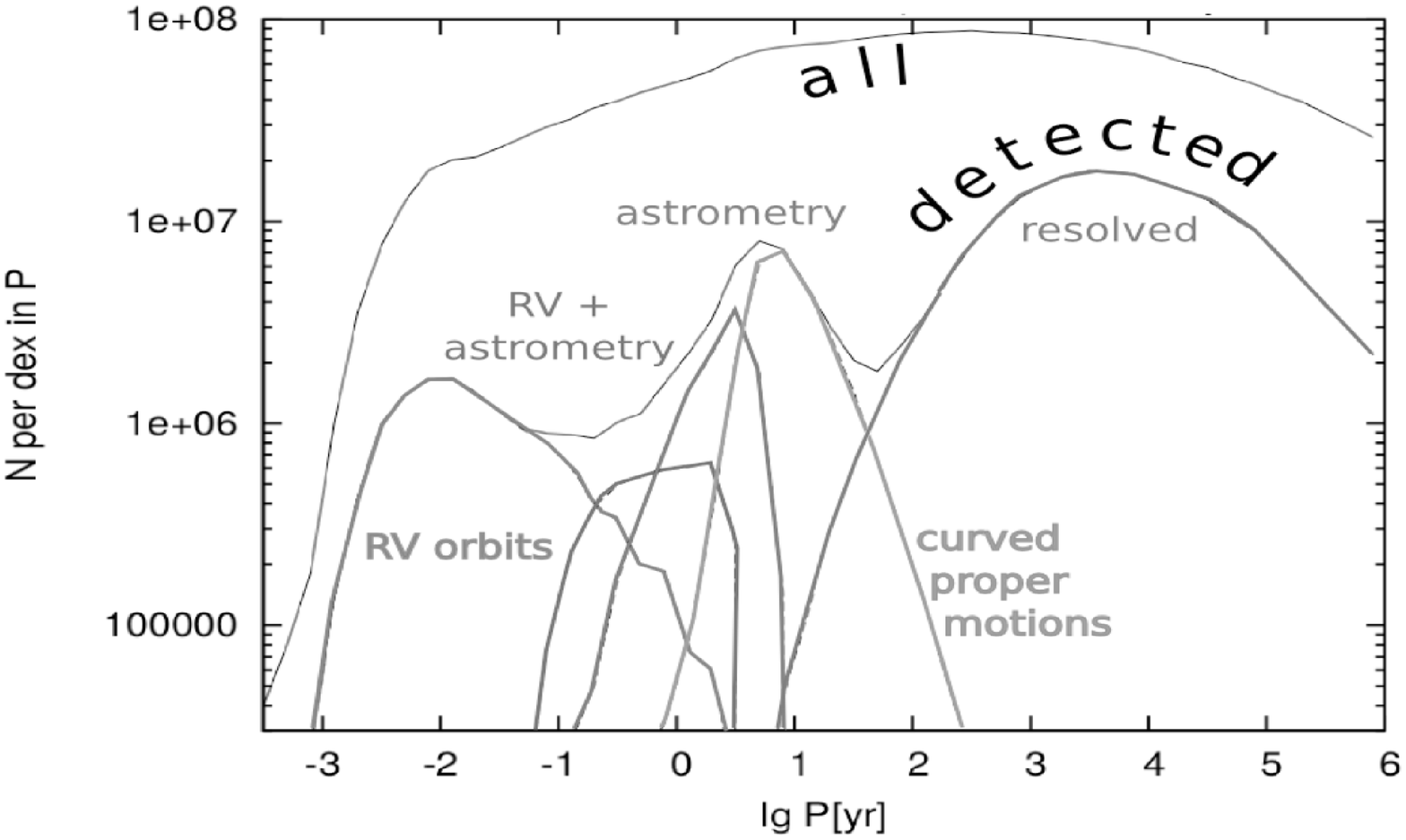}{figBinaries}{
Expected binary census of Gaia for different types of observing techniques. Adapted from \citet{Soderhjelm2004}.
}

Gaia will observe huge numbers of different types of binaries \citep{Zwitter2004,Eyer2015} and study them with a wide range of techniques (Fig.\ \ref{figBinaries}). One of its core strengths will be a derivation of accurate astrometric solutions even for binaries with extreme mass ratios \citep{Soderhjelm2004}. On the other hand spectroscopy from ground based surveys will be the source of detailed chemistry for any type of binary or multiple system. Astrometry has been frequently used to supplement spectroscopic observations in the past \citep[e.g.][]{Sahlmann2013}, but in Gaia the opposite will be a common case \citep[e.g.\ ][]{Torres2006}. Many astrometric binaries will have components of similar mass and luminosity. The reach of astrometry is limited in this case: the two stellar images are usually not spatially resolved, so that Gaia will be able to trace only the astrometric motion of the photocenter of the two components. Such studies yield accurate orbital period, but since the photocenter is located somewhere between the two stars individual masses cannot be derived from astrometry alone. Here even a single spectrum obtained during a spectroscopic survey can be extremely valuable. Radial velocities of individual components in an SB2 at an orbital phase known from astrometry allow to derive the true sizes of both orbits, and so the complete solution of the system. A proper Bayesian analysis of simultaneous astrometric and spectroscopic information will be needed for this task \citep{Schulze2012}.




\end{document}